\documentclass[preprint,twocolumn,5p]{elsarticle}
\usepackage{graphicx}
\usepackage[dvips]{color}
\expandafter\let\csname equation*\endcsname\relax
\expandafter\let\csname endequation*\endcsname\relax
\usepackage{amsmath}
\usepackage[normal]{subfigure}
\usepackage[english]{babel}

\journal{Chaos, Solitons and Fractals}

\begin{document}

\begin{frontmatter}
\title{Anomalous transmission and drifts in one-dimensional L\'evy structures}

\author[1]{P. Bernab\'o}
\author[2]{R.  Burioni}
\author[3,4]{S. Lepri}
\author[2,5]{A. Vezzani}

\address[1]{Universit\`a degli Studi di
Pisa, Dipartimento di Ingegneria dell'Informazione, Via G. Caruso 16
56122 - Pisa,  Italy
}

\address[2]{Universit\`a degli Studi di
Parma, Dipartimento di Fisica, viale G.P.Usberti 7/A, 43100 Parma, Italy
}

\address[3]{Consiglio Nazionale
delle Ricerche, Istituto dei Sistemi Complessi,
via Madonna del Piano 10, I-50019 Sesto Fiorentino, Italy
}

\address[4]{Istituto Nazionale di Fisica Nucleare, Sezione di Firenze,
via G. Sansone 1 I-50019, Sesto Fiorentino, Italy}

\address[5]{Consiglio Nazionale
delle Ricerche, Centro S3, Istituto di Nanoscienze,
Via Campi 213A, I-41125 Modena Italy}

\begin{abstract}
We study the transmission of random walkers through a finite-size
inhomogeneous material with a quenched, long-range correlated
distribution of scatterers. We focus on a finite one-dimensional structure where walkers undergo random collisions with a subset of sites distributed on deterministic (Cantor-like) or random positions, with L\'evy spaced distances. Using scaling arguments, we consider stationary and time-dependent transmission
and we provide predictions on the scaling behavior of particle current as a function of the sample size.
We show that, even in absence of bias, for each single realization a non-zero drift can be present, due to the intrinsic asymmetry of each specific  arrangement of the scattering sites. For finite systems,
this average drift is particulary important for characterizing the transmission properties of individual samples.
The predictions are tested against the numerical solution of the associated
master equation. A comparison of different boundary conditions is given.
\end{abstract}


\end{frontmatter}

\section{Introduction}

Transport and diffusion in complex systems is often
anomalous, since the  basic hypotheses underlying the laws of ordinary Brownian motion can
be violated. Specific examples are abundant in physics and chemistry, ranging from porous
media, to tracer motion in turbulent fluids and plasmas
\cite{AnotransBook08}. Interdisciplinary applications of
non-Brownian processes also arised recently in diverse fields
as animal movement \cite{Smouse2010} and social and cognitive
phenomena \cite{Baronchelli2013}. In all these cases, the lengths of the steps taken by the diffusing particles can have large fluctuations, and typically follow a probability distribution with heavy tails.

Among the many possible experimental applications in physics,
of special interest is the recent realization of materials termed L\'evy glasses, where
light rays propagate through an assembly of transparent spheres embedded
in a scattering medium \cite{Barthelemy2008,Bertolotti2010b}. If the diameter
of spheres is designed to have a power-law distribution, light can indeed perform anomalous
diffusion. From the theoretist viewpoint,  a salient feature of such experiment is that the
spatial arrangement of the scattering media is fixed for each sample,
i.e. the disorder is quenched. This implies that the walk is correlated as
light that has just crossed a large glass sphere
has a larger probability of being backscattered at the following step
and thus to perform a jump of roughly the same length.

The reference model for this class of
phenomena is the so-called L\'evy walk \cite{Blumen1989,Klafter1990,Geisel1995}, in which particles
perform independent steps $l$ at constant velocity, with a  distribution
following an algebraic tail of the form $l^{-(1+\alpha)}$ for large $l$.
Such a heavy-tailed distribution arises in presence of
dynamical correlations, like in the
case of diffusion in chaotic and intermittent systems \cite{Geisel1985,Geisel1995},
or from complex interaction with the environment \cite{Cipriani05,Delfini07a}.
When the variance of $l$ diverges, for
$\alpha< 2$, transport is thereby dominated by very long steps, the mean square displacement
increases faster than linearly  with time, and transport is superdiffusive.

While the case of uncorrelated jumps (annealed) is well understood,
quenching effects are known to affect strongly the diffusion
properties \cite{Kutner1998,Fogedby94,Schulz02} and in this case there are still many open
problems. Relevant features of experiments on scattering
in inhomogeneous media can be described as a random walk in a quenched,
long-range correlated environments
and can be studied directly by Monte Carlo simulations \cite{Barthelemy2010,Groth2012}.
However, simplified models are of great help to provide theoretical insight.
In this context, a minimal model that includes the effects of disorder and
anomalous diffusion consists of a free particle moving
through a one-dimensional array of scatterers whose spacing is power-law distributed
\cite{Barkai2000,Beenakker2009,Burioni2010}. In this spirit, a closely related
class of self similar models, termed Cantor
graphs has been also considered \cite{Burioni2010a}.
As the latter is generated by deterministic rules, diffusion properties
can be investigated analytically using
tools from random walk theory on directed graphs \cite{Burioni2010a}.
In both cases, random walks through such structures are naturally
correlated, due to the long jumps induced by the underlying self similar topology.

The present work aims at understanding transport in this class of finite systems.
This is a relevant issue for the interpretation of experiments, that typically
deal with transport through finite samples (e.g. slabs) \cite{Barthelemy2008}.
In the framework of random walk theory
much progress has been made in the last years in the characterization of
superdiffusive motion in infinite domains  \cite{Metzler2004,AnotransBook08}.
On the other hand, the case of finite systems is relatively less developed
\cite{Davis1997,Drysdale1998,Larralde1998,Buldyrev2001,vanMilligen08,Lepri2011,Dhar2013}.
The focus here will be on both stationary transport and time-dependent
transmission for finite, one-dimensional structures. The study will be
undertaken by means of a formulation in terms of a master equation for
the process. This has several advantages for the theoretical analysis,
since it deals directly with probabilities instead of ensembles
of individual trajectories.

For quenched disorder, it is known that different
averaging procedures \cite{Bouchaud90} leads to different results. For the present class of models
this requires for instance to distinguish among the possible types of
initial conditions \cite{Barkai2000,Burioni2010a}.
In general, the averaging over an ensemble of trajectories still
depends on the realization of the disorder. In particular, despite
the walker is unbiased, for each single realization there may
be a non-zero drift due to the intrinsic asymmetry of each specific
arrangement of the scattering sites. For finite systems,
this average drift is particulary important for
characterizing the transmission properties of individual samples.
Indeed, there may be cases in which the drift motion over the
observational time is of the same order of the average diffusive
spreading, thus affecting substantially the measurements.
In the second part of the paper we will address this question
by examining the scaling behavior of such drift and its statistics.

The paper is organized as follows. In section \ref{sec:model} we recall the
definition of the model and its dynamics. In Section \ref{sec:stationary}
we discuss the transport properties of a finite lattice under stationary
conditions, i.e. when a constant flux of particles is kept at one side
of the system. In this case, the models allows for analytic calculations.
Then, in section \ref{sec:timed} we turn to the
time-resolved problem, namely to the situation in which particles
are injected only at initial time at one lattice boundary.
The related issue of the average drift induced by the statistical
fluctuations of the structure in the random case is
addressed in section \ref{sec:drift}.

\section{The model}
\label{sec:model}

We consider a discrete-time random walk on a one-dimensional lattice.
When the walker arrives at the $n$th site it can be transmitted with
probability $T_n$ (and reflected with probability $R_n=1-T_n$.
Clearly,  the speed is conserved during
the evolution and we can set its value to unity without loss of generality.
Two types of sites are represented, the ``transparent''
ones for which $T=1$ that correspond to a completely ballistic propagation
and the ``scattering'' ones, where $T_n=T$, ($0<T<1$). In the following
we set $T=1/2$, as  the results are expected not to be
affected by this choice up to inessential prefactors \cite{Burioni2012}.

\subsection{The Master Equation}

The master equation of a random walker on one-dimensional lattice, accordingly with
Persistent Random Walk model, is
\begin{eqnarray}
\label{Master_Equation_p+}
p_{n}^{+}(t+1)=T_{n-1}p_{n-1}^{+}(t)+R_{n-1}p_{n-1}^{-}(t)\\
\label{Master_Equation_p-}
p_{n}^{-}(t+1)=R_{n+1}p_{n+1}^{+}(t)+T_{n+1}p_{n+1}^{-}(t)
\label{Master_Equation}
\end{eqnarray}
where $p^{\pm}_{n}$ is the walker's probability to land at site $n$ with positive velocity from the left side ($+$),
or with negative velocity from the right side ($-$), so the total probability is $p_{n}(t)=p_{n}^{+}(t)+p_{n}^{-}(t)$.
In this formulation, our model can be regarded as a random walk
with a site-dependent persistence (see e.g. \cite{Miri2006} and
references therein).

\subsection{The random and deterministic correlated L\'evy Structures}

\begin{figure}[htbp]
\centering
 {\includegraphics[width=0.45\textwidth,clip]{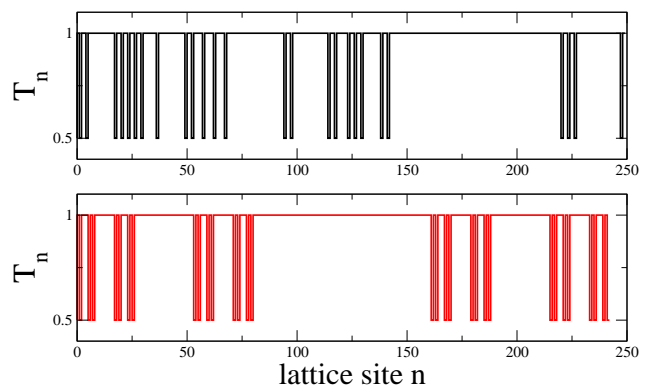}}
 \caption{\protect\footnotesize Illustration of the structures: transmission coefficient profiles
 $T_n$ for the random (upper panel) and Cantor (lower panel) cases; $\alpha=0.630930$ corresponding
 to $n_r=2$, $n_u=3$.}
\label{schema}
\end{figure}

We model correlated media  in which the scatterers are distributed
on a self similar set, such that the regions of ballistic sites are distributed according to
a fat-tail. We will consider two classes of models (see fig.\ref{schema}).
The first is random and has been first introduced in \cite{Barkai2000}
(see also \cite{Beenakker2009} and \cite{Burioni2010a}).
Here, the probability $\lambda(r)$ to have two consecutive scattering sites
separated by $r$ ballistic ones is
\begin{equation}
  \label{p_r}
  \lambda(r)\equiv \frac{a}{r^{1+\alpha}},  ,
\end{equation}
where $\alpha>0$, $r$ is a positive integer and $a$ a suitable normalization constant.
Different realizations of the structure can be easily generated via a
non-uniform random variate algorithm described in \cite{Devroye1986}.

The second type is a class of deterministic quasi-lattices (see again fig.\ref{schema}),
built by placing the scatterers on generalized Cantor
sets \cite{Burioni2010}. Each sample, and the ensuing step length distribution,
is defined by two parameters, $n_u$ and $n_r$, used in its recursive construction.
The former represents the growth of
the longest step when the structures is increased by a generation, so that the longest
step in a structure of generation $G$ is proportional to $n_u^G$; $n_r$ is the number
of copies of generation $G-1$ that form the generation $G$, so that the total number
of scatterers in the generation $G$ is proportional to $n_r^{G}$ (see \cite{Burioni2010}
for details). For this class of structures,
the role of the exponent $\alpha$ of the random case
is played by $\alpha = \log n_r /  \log n_u$ \cite{Burioni2010}.
Moreover, it can be shown that if  $n_u< n_r$ (i.e. $\alpha>1$)
the fraction of scattering sites remains strictly positive in the thermodynamic
limit $G\to \infty$. Thus we will refer to such a case as to the so
called \emph{fat} fractals. An example of this kind of structure is the
Smith-Volterra-Cantor set. Conversely,
we will name \emph{slim} fractals the graphs with
$n_u > n_r$ (i.e. $\alpha<1$), where the measure of the scattering
region is vanishing.

\subsection{The Boundary conditions}

In this work we are interested in finite lattices of a fixed number of sites, labeled by the
integer $n$ with $n=1,\ldots ,N$. We thus need
to specify the boundary conditions, by fixing the values of the probabilities
on the $0$th site on the left side ($L$) and $N+1$th site on the right side ($R$).
Let us define the probabilities on these sites $p_{0}^{\pm}=p_{L}^{\pm}$ and $p_{N+1}^{\pm}=p_{R}^{\pm}$, so that:
\begin{equation}\label{probabilita totali L-R}
p_{L}=p_{L}^{+}+p_{L}^{-}, \qquad p_{R}=p_{R}^{+}+p_{R}^{-}
\end{equation}
We also need to specify the currents at the system boundaries. To this aim
let us consider the total probability
\begin{equation}\label{probabilita totale}
p_{TOT}(t)=\sum_{n=1}^{N}\left[p^{+}_{n}(t)+p^{-}_{n}(t)\right]
\end{equation}
and using the master equation (\ref{Master_Equation}) we can write the continuity equation
\begin{equation}\label{equazione di continuita}
p_{TOT}(t+1)=p_{TOT}(t)+J_{L}+J_{R}\;,
\end{equation}
where we define
\begin{eqnarray}
\label{flusso left}
&& J_{L}   \equiv 
R_{0}p^{-}_{0} +T_{0}p^{+}_{0}-R_{1}p^{+}_{1}-T_{1}p^{-}_{1} \\
\label{flusso right}
&& J_{R} \equiv 
R_{N+1}p^{+}_{N+1}+T_{N+1}p^{-}_{N+1}-R_{N}p^{-}_{N}-T_{N}p^{+}_{N}\;\;.
\end{eqnarray}

%

In the following we will compare two different cases:

\textit{Mixed boundary conditions} where we impose a total reflection
on the left side and an absorbing condition on the right,
\begin{equation}\label{mbc left}
J_{L}=0, \quad p^{+}_{0}=0, \quad p^{+}_{N+1}=0, \quad p^{-}_{N+1}=0.
\end{equation}
Substituting (\ref{mbc left}) in (\ref{flusso left}) and (\ref{flusso right}) we obtain
\begin{equation}\label{condizione mbc}
R_{0}p^{-}_{0}=R_{1}p^{+}_{1}+T_{1}p^{-}_{1}
\end{equation}
and the flux on the right side is
\begin{equation}\label{flusso right mbc}
J_{R}=-R_{N}p^{-}_{N}-T_{N}p^{+}_{N}\;.
\end{equation}

\textit{Absorbing boundary conditions},  defined as
\begin{equation}\label{abc left}
p^{-}_{0}=0,\quad p^{+}_{0}=0, \quad
p^{+}_{N+1}=0, \quad p^{-}_{N+1}=0\;.
\end{equation}
In this case, equations (\ref{flusso left}) e (\ref{flusso right}) reads:
\begin{equation}\label{flusso left abc}
J_{L}=-R_{1}p^{+}_{1}-T_{1}p^{-}_{1}
\end{equation}
and
\begin{equation}\label{flusso right abc}
J_{R}=-R_{N}p^{-}_{N}-T_{N}p^{+}_{N}\;\;.
\end{equation}

\section{The Stationary solution}
\label{sec:stationary}

Let us consider the time-independent solution of (\ref{Master_Equation}) on a finite lattice.
It can be shown that such solution is obtained by transfer matrix method
\begin{equation}\label{formamatricialeME}
\begin{pmatrix}
p^{+}_{n+1}\\
p^{-}_{n+1}
\end{pmatrix}=
\begin{pmatrix}
T_{n} & R_{n} \\
-\frac{R_{n+1}T_{n}}{T_{n+1}} & \frac{1-R_{n}R_{n+1}}{T_{n+1}}
\end{pmatrix}
\begin{pmatrix}
p^{+}_{n}\\
p^{-}_{n}
\end{pmatrix}\;\;\;,
\end{equation}
and using boundary conditions, we can write the general form of equation (\ref{formamatricialeME})
\begin{equation}\label{forma matriciale equazione trasferimento L-R}
\begin{pmatrix}
p^{+}_{R}\\
p^{-}_{R}
\end{pmatrix}=
\begin{pmatrix}
m_{11} & m_{12}\\
m_{21} & m_{22}
\end{pmatrix}
\begin{pmatrix}
p^{+}_{L}\\
p^{-}_{L}
\end{pmatrix}=
M
\begin{pmatrix}
p^{+}_{L}\\
p^{-}_{L}
\end{pmatrix}\;\;\;,
\end{equation}
where the $2\times2$ matrix $M$ is the product of $N+1$ transfer matrixes,
Using equation (\ref{probabilita totali L-R}) we have also
\begin{equation}\label{flusso L-R}
J_{L}=p_{L}^{+}-p_{L}^{-},\qquad J_{R}=p_{R}^{+}-p_{R}^{-}
\end{equation}
and it is possible to evaluate the output flux $J_{R}=J_{L}=J$.

The matrix $M$ is computed conveniently distinguishing between
scattering and transparent sites. Obviously, for ballistic sites the associated matrix is the identity.
For a block of $s$ consecutive scattering sites the transfer matrix is
\begin{eqnarray}
B_{s} &=&
\begin{pmatrix}
T & R\\
0 & 1
\end{pmatrix}
\begin{pmatrix}
T & R\\
-R & 2-T
\end{pmatrix}^{s-1}
\begin{pmatrix}
1 & 0\\
-\frac{R}{T} & \frac{1}{T}
\end{pmatrix}= \nonumber \\
&=& \begin{pmatrix}
1-s\frac{R}{T} & s\frac{R}{T}\\
-s\frac{R}{T} & 1+s\frac{R}{T}
\end{pmatrix}\quad,
\label{matrice Bs}
\end{eqnarray}
where we have used the matrix identity ($n$ positive integer)
\begin{equation}\label{proprieta matrici}
\begin{pmatrix}
T & R\\
-R & 1+R
\end{pmatrix}^{n}=
\begin{pmatrix}
1-nR & nR\\
-nR & 1+nR
\end{pmatrix}
\end{equation}
If there are $m$ equal blocks in the lattice $1...N$, the total transfer matrix is the product of $m$ times $B_{s}$,
namely
\begin{equation}
M = B^{m}_{s}=
\begin{pmatrix}
1-ms\frac{R}{T} & ms\frac{R}{T}\\
-ms\frac{R}{T} & 1+ms\frac{R}{T}
\end{pmatrix}\;\;\;.
\end{equation}

For the Cantor model at generation $G$ the number of consecutive scatterers $N_{b}=2n_{r}^{G}$,
where $n_{r}$ is the number of replicas.
where $ms=2n_{r}^{G}$.
Resolving the system and using (\ref{probabilita totali L-R}), (\ref{flusso L-R}) we obtain
\begin{equation}
J=-\frac{T}{4n_{r}^{G}R}\left( p_{R}-p_{L}\right)=-\frac{T}{4n_{r}^{G}R}\frac{\Delta p}{\Delta x}N
\end{equation}
where $\frac{\Delta p}{\Delta x}$ is the one-dimensional probability gradient. Using the first Fick's law we can write
\begin{equation}
D=\frac{N}{4n_{r}^{G}\frac{R}{T}}\;\;\;.
\end{equation}

For the random model we assume that there are $m$ subsystem composed by only one ($s=1$) scattering site separated by a random distance $l$ with a
power-law distribution $l^{-1-\alpha}$.
Similarly to the Cantor model, using (\ref{probabilita totali L-R}) e (\ref{flusso L-R}) we obtain
\begin{equation}
J=-\frac{T}{2mR}\left( p_{R}-p_{L}\right)=-\frac{T}{2mR}\frac{\Delta p}{\Delta x}N
\end{equation}
and the diffusion constant
\begin{equation}
D=\frac{N}{2m\frac{R}{T}}\;\;\;.
\end{equation}
The problem is thus to determine the statistics of the random variable  $m$ for fixed $N$.
This is a nontrivial problem \cite{Beenakker2009}. A simple estimate can be obtained as
follows. Since $N/m$ is the average distance between two consecutive scatterers we can write
\begin{equation}
\frac{N}{m}=\sum_{l=0}^{N}l^{-\alpha}\simeq\int_{0}^{N}l^{-\alpha}dl=\frac{N^{1-\alpha}}{1-\alpha}\;\;\;,
\end{equation}
and so we obtain
\begin{equation}
D=\frac{N^{1-\alpha}T}{2|1-\alpha|R}\;\;\;,
\label{coefficiente diffusione caso staz rnd}
\end{equation}
Notice that the diffusion constant is proportional to $T/R$ and it is finite for
$\alpha>1$ and diverging for $\alpha<1$. This is consistent with the results
for the infinite domain \cite{Burioni2010,Burioni2010a}.

\section{Time-resolved transmission}
\label{sec:timed}

Let us now turn to the case of time-resolved transmission. The type of experiment
we have in mind is the following: an input pulse is applied at one boundary of the
system and the output at the other side is observed. Some relevant prediction
can be obtained by scaling arguments. The main quantity to consider is the probability for a
walker to be at time $t$ at distance $r$ from the starting point, which we
denote by $p(r,t)$. To investigate it dependence sample size, we consider
its dynamical scaling properties.

The dynamical scaling hypothesis amounts to the fact that, on
infinite domains, the evolution mostly depends on a single scaling length $\ell(t)$.
More precisely \cite{Burioni2010,Burioni2010a}
$p(r,t)$ can be written in scaling form as:
\begin{equation}
p(r,t)=\frac{1}{\ell (t)} f\left(\frac{r}{\ell(t)}\right),
\label{scalp}
\end{equation}
for the random model and as:
\begin{equation}
p(r,t)=\frac{1}{\ell (t)} f\left[ \frac{r}{\ell(t)},g\left[ \log_{n_u}\ell(t)\right]\right]
\end{equation}
for the deterministic Cantor samples, with $g$ periodic function (of unit period) accounting for
log-periodic oscillations on the deterministic self-similar structures \cite{Burioni2010a}.
In the random case, the scaling function (\ref{scalp})
can also present a subleading (i.e. vanishing in probability) term that can influence the
evaluation of high order moments \cite{Burioni2010}.
The scaling information is actually given by the growth law of the scaling length, which
is predicted to be:
\begin{equation}
\ell(t) \;\sim\;
\begin{cases}
t^{\frac{1}{1+\alpha}} & \mathrm{if}\ 0<\alpha<1 \\
t^{1 \over 2} & \mathrm{if}\ 1 \leq \alpha
\label{ellcom}
\end{cases}
\end{equation}
%
%

Let us now turn to finite lattices of $N$ sites, and estimate the size dependence from the scaling form. For walkers starting at time $t=0$ from the left border we
consider the time-resolved transmitted current $J_{R}(t,N)$. Using the above scaling hypotheses
one would argue that, at leading order, for $\alpha<1$, $J_R(t,N)=B(N) G (t/N^{1+\alpha})$ where $G(\cdot)$  is a scaling
function. The coefficient $B$ can be estimated as follows. The time-integrated current $\int_0^\infty J_R(t,L) dt$
is by definition the total number of walkers escaping from the rightmost boundary. This, in turn, must be
proportional to the conductivity,  i.e. to $N^{-\alpha}$ in the case of absorbing boundary conditions, and to $1$ for the mixed ones (all the particles exit from the rightmost side in this case). A straightforward
calculation thus yields
\begin{equation}\label{scaling flux MBC}
J_{R}(t,N)=\frac{1}{N^{1+\alpha}}G\left( \frac{t}{N^{1+\alpha}}\right)\;\;\;,
\end{equation}
for mixed boundary conditions, and
\begin{equation}\label{scaling flux ABC}
J_{R}(t,N)=\frac{1}{N^{1+2\alpha}}G\left( \frac{t}{N^{1+\alpha}}\right)
\end{equation}
for absorbing conditions. For the case $\alpha>1$ a similar reasoning leads to the result
\begin{equation}\label{scaling flux MBC 2}
J_{R}(t,N)=\frac{1}{N^{2}}G\left( \frac{t}{N^{2}}\right)\;\;\;,
\end{equation}
for mixed boundary conditions, and
\begin{equation}\label{scaling flux ABC 2}
J_{R}(t,N)=\frac{1}{N^{3}}G\left( \frac{t}{N^{2}}\right)
\end{equation}
for absorbing conditions.

\begin{figure}[htbp]
\centering
 \subfigure[]
   {\includegraphics[width=0.32\textwidth]{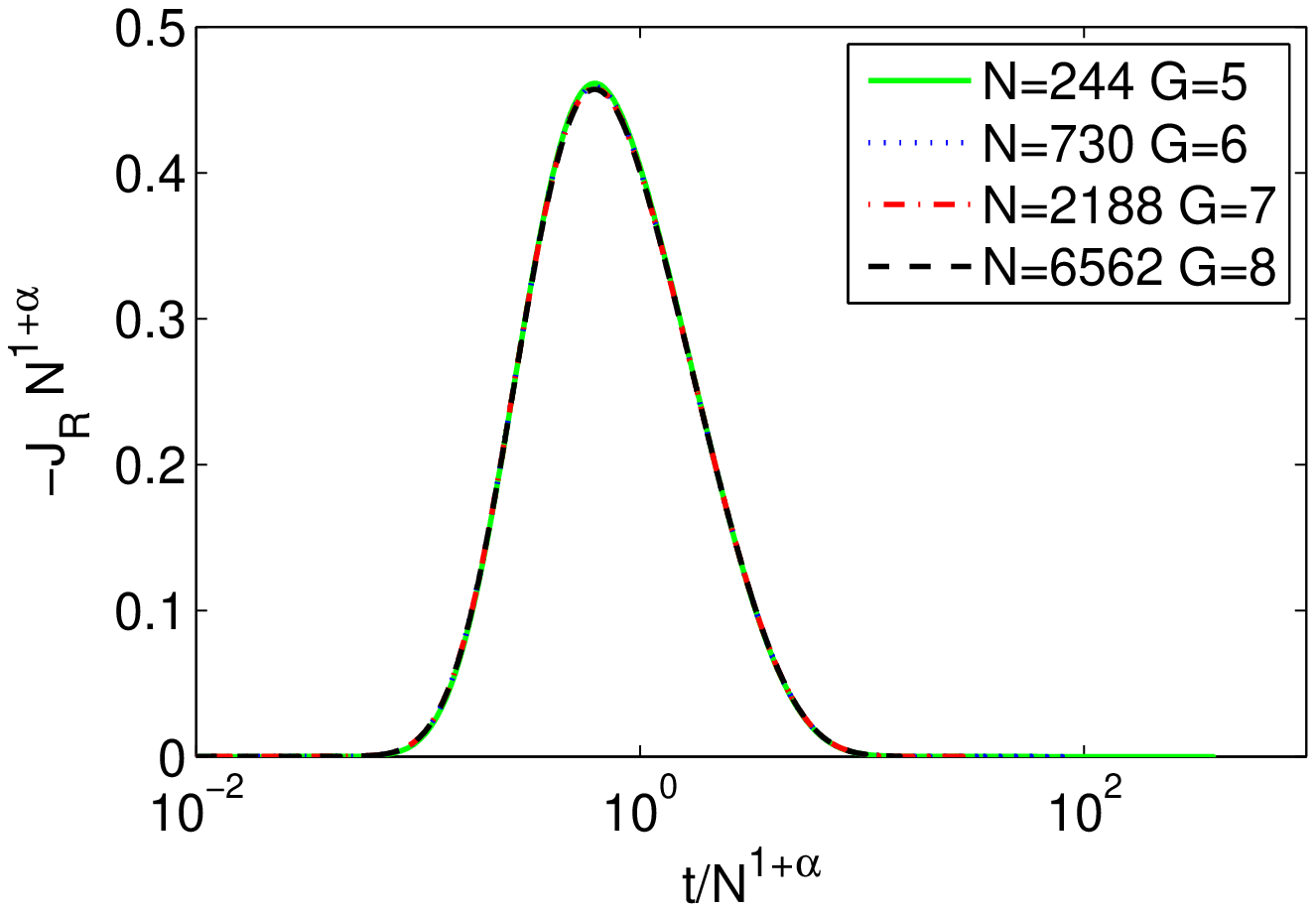}}
 \subfigure[]
   {\includegraphics[width=0.32\textwidth]{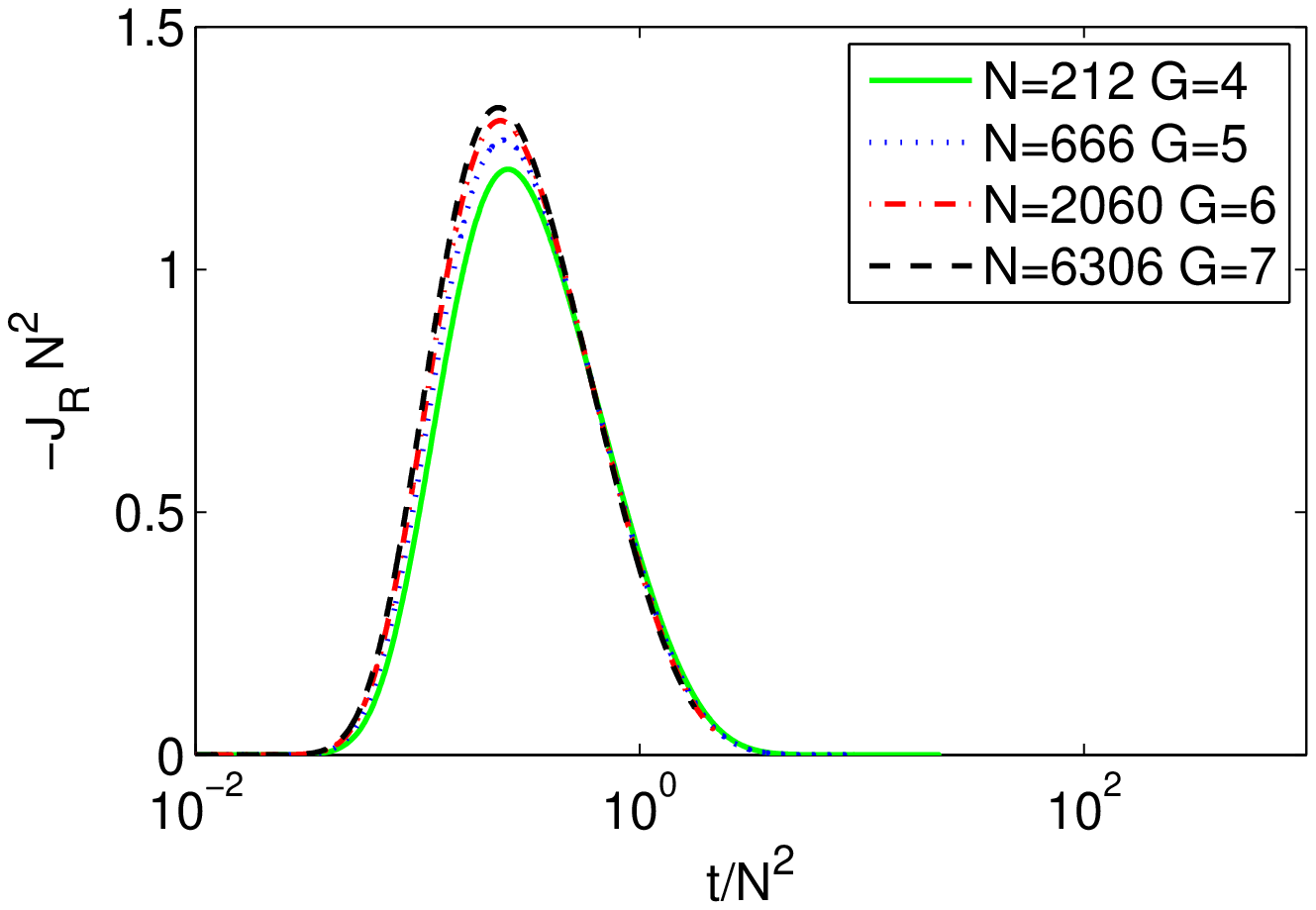}}
 \subfigure[]
   {\includegraphics[width=0.32\textwidth]{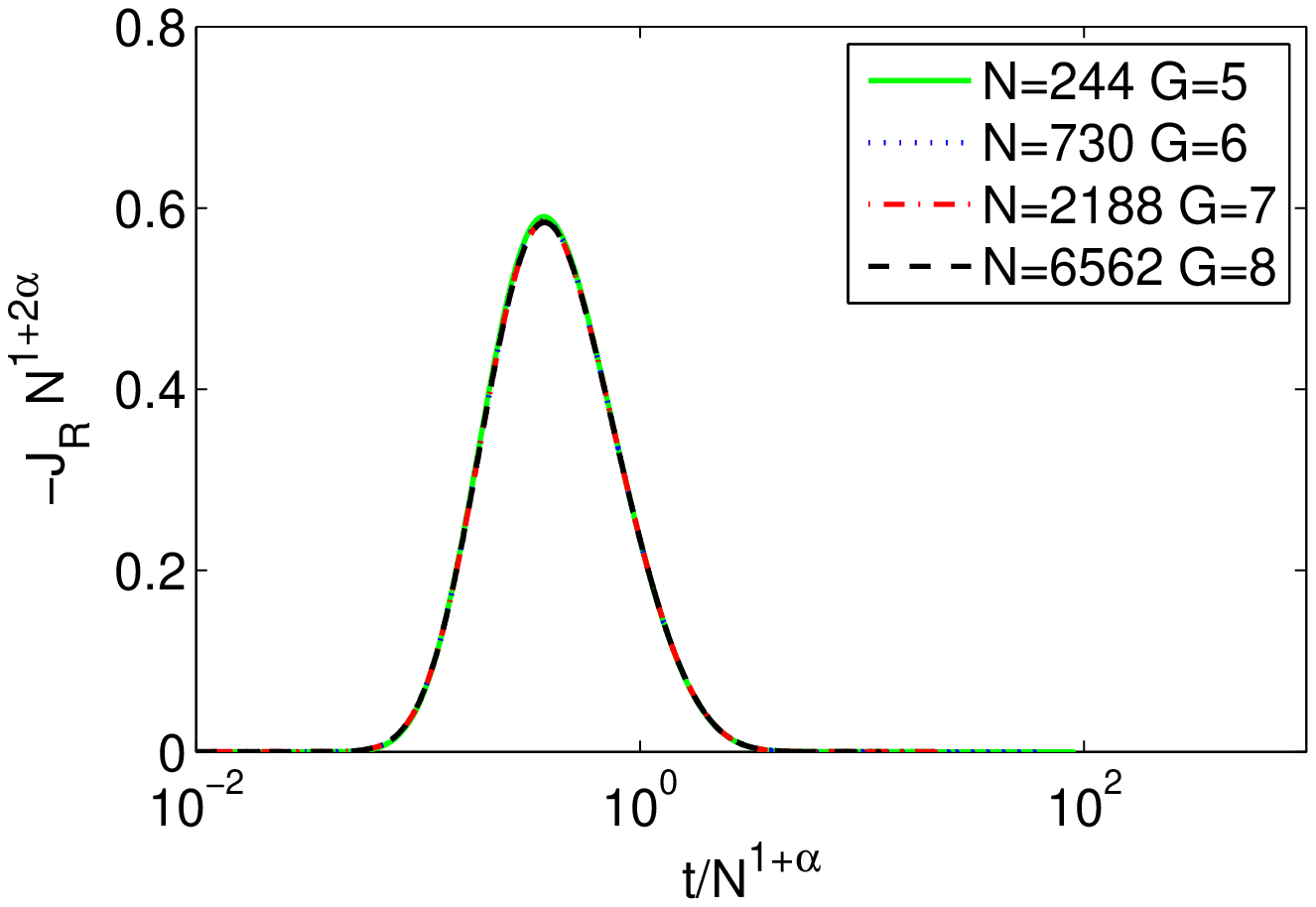}}
 \subfigure[]
   {\includegraphics[width=0.32\textwidth]{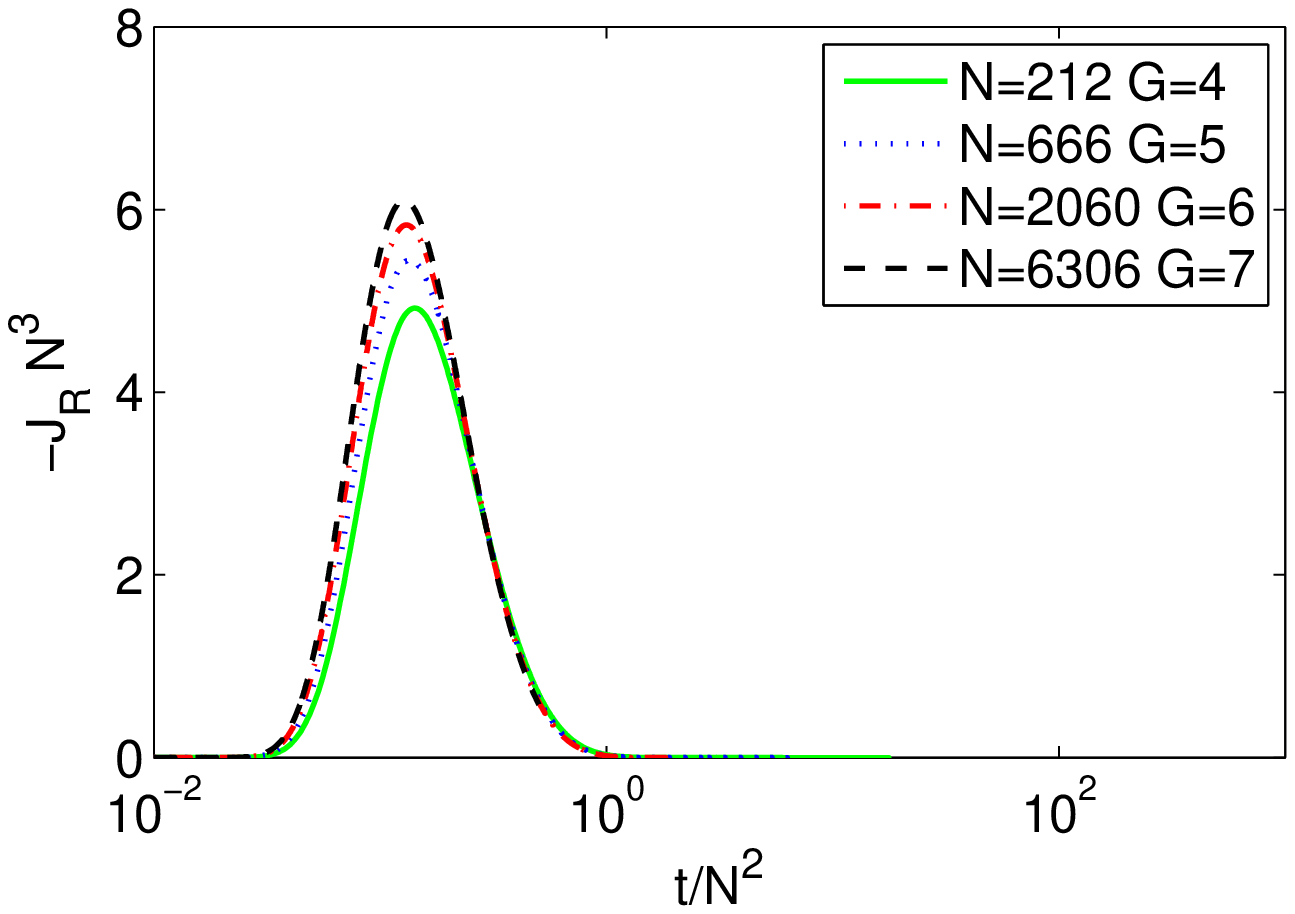}}
 \caption{\protect\footnotesize Cantor model: dynamic scaling of $J_{R}(t,N)$ for different system sizes $N$: slim Cantor structure (a) and (c) ($n_{r}=2$, $n_{u}=3$ corresponding to $\alpha=0.630930$) and fat Cantor structure (b) and (d) ($n_{r}=3$, $n_{u}=2$ corresponding to $\alpha=1.584963$) with mixed boundary conditions and
 absorbing conditions respectively.}
\label{scalingfluxcantor}
\end{figure}

\begin{figure}[htbp]
\centering
 \subfigure[]
   {\includegraphics[width=0.35\textwidth ]{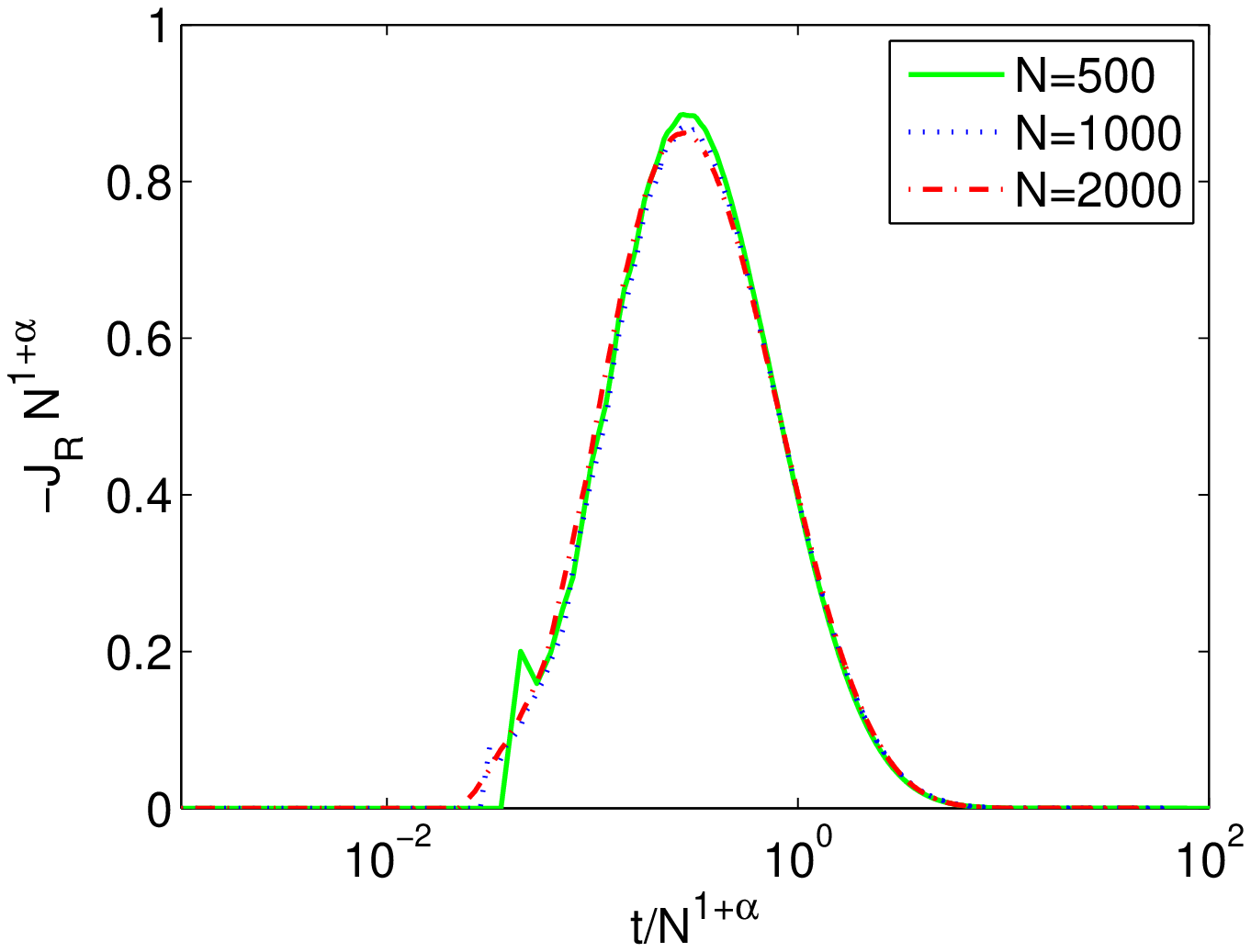}}
 \hspace{5mm}
 \subfigure[]
   {\includegraphics[width=0.35\textwidth ]{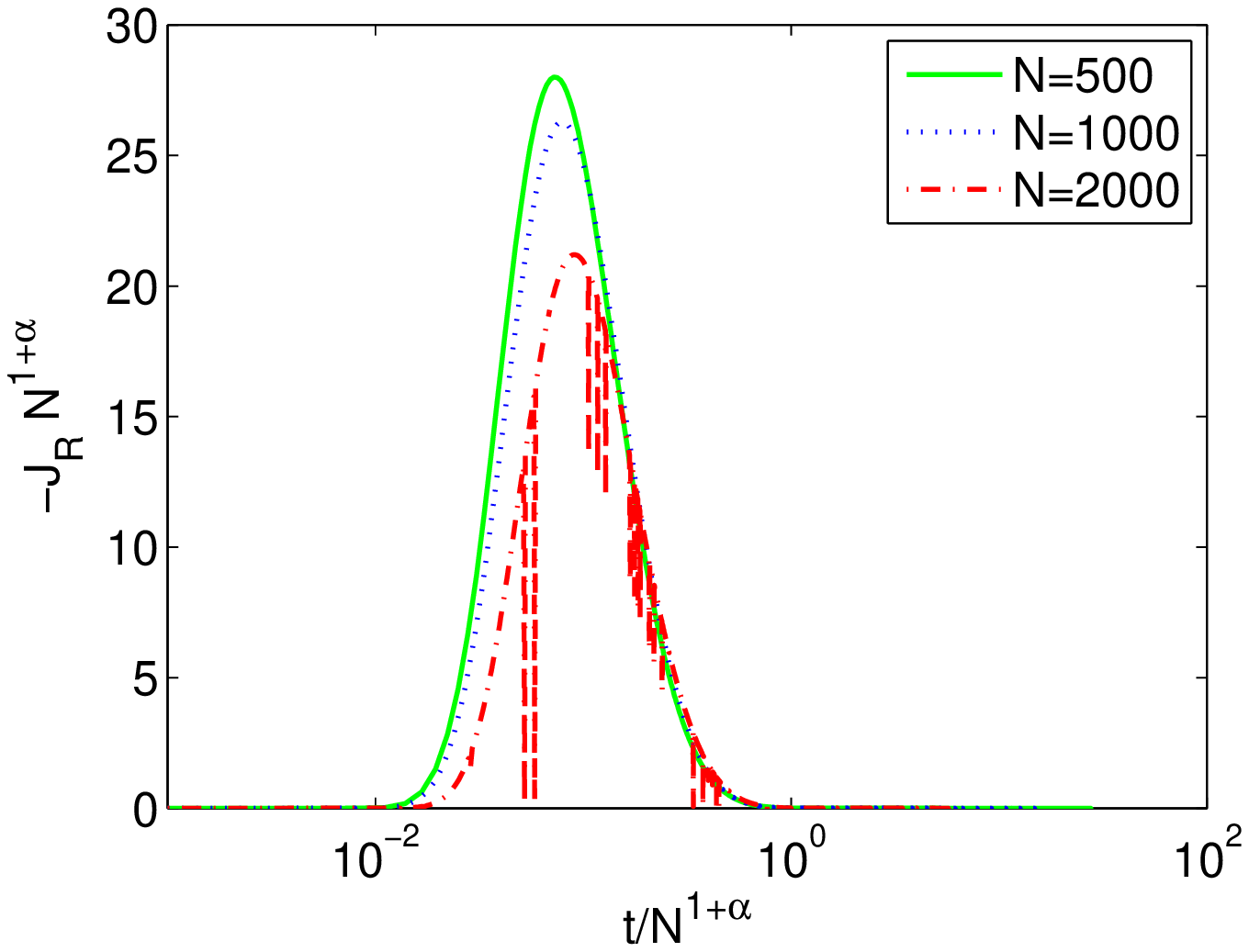}}
 \caption{\protect\footnotesize Random model: dynamic scaling of $J_{R}(t,N)$
 with mixed boundary conditions averaged on about $500$ realizations of disorder and $\alpha=0.5$:
 in fig. (a) the average of $J_{R}(t)$ is obtained with all random configurations, on the other hand in
 fig. (b) the average is obtained removing the configurations having no or very few scattering sites
 (see text).}
\label{scalingfluxrndmbc}
\end{figure}

The above prediction have been tested by solving iteratively the master equation (\ref{Master_Equation})
for lattices of different sizes $N$ and
for both mixed and absorbing boundary conditions. The initial condition
is impulsive on the first site of the chain, $p^{\pm}_{n,1}(0)=\frac{\delta_{1,n}}{2}$.
In fig.~\ref{scalingfluxcantor} and \ref{scalingfluxrndmbc} we report the results for Cantor and
the random model for different boundary conditions. An issue here concerns the
sampling of disorder realizations. In the process of generating such realization it is often
the case that some of them are devoided of scattering sites (except for the $n=1$ one that we
always fix to have $T_1=1/2$). Such realizations will affect the ensemble-averages via
large ballistic peaks that hinder a meaningful comparison with the expected theoretical
estimates. We thus decided to cutoff the distribution $\lambda$ up to some $r_{max}\sim N$,
meaning that we consider only realizations where there is a minimal number of scatterers.
This anyhow ensures that the scaling test is significant as $N\to \infty$.  As seen in
fig.~\ref{scalingfluxrndmbc} the data averaged of
such ensemble nicely obey the expected scaling for both types of boundary conditions.

\section{The average drift and its fluctuations in the random model}
\label{sec:drift}

The class of system we are dealing with are characterized by a quenched disorder,
so that the local transmission rates are fixed and are independent of time.
Despite that the walker is unbiased, for each single realization there may
be a nonzero average displacement, due to the intrinsic asymmetry of each specific
realization of the system (see for instance the upper panel of fig.~\ref{schema}).
In this section we aim at characterizing such a displacement and its
sample-to-sample fluctuations. The main observables of interest we considered
are the moments:
\begin{equation}
x_q=\left\langle\; \left| \sum_{n=1}^{N} (n-n_{0})\, p_{n}(t) \right|^q \;\right \rangle\;\;\;= \langle |x_B|^q \rangle
\label{momentidef}
\end{equation}
Notice that the definition implies an average over trajectories and a
further average over realization of the random structure, denoted by
$\langle \ldots \rangle$. The variable
$x_B= \sum_{n=1}^{N} (n-n_{0})\, p_{n}(t)$ denotes the average position of the walker
or the baricenter of the probability
distribution, in a single disorder realization.
Of course, as the structures are statistically symmetric under spatial
reflection, the absolute value in the definition (\ref{momentidef})
is crucial to yield a nonzero ensemble average.
The moments are assumed to grow asymptotically as $x_q \sim t^{\gamma(q)}$.
The fact that $\gamma(q)$ will be a nontrivial function of $q$ it is
a signature of the so-called strongly anomalous diffusion \cite{Castiglione1999}.

The behavior of the moments of $x_B$ can be inferred as follows. In a disorder realization,
let us consider the first $n$ scattering sites placed
to the right and to the left with respect $x=0$, that denotes the scatterer corresponding to
the starting point of the random walk. The positions of the
$n$ scatterers are
$X_{\pm n}=\pm \sum_{j=1}^n r_j^\pm$
where $r_j^\pm$ are integers
extracted from the spacing distributions $\lambda(r)$. The baricenter $B$ of the
region between $X_{n}$ and $X_{- n}$ is therefore
$B=\frac {X_n-X_{- n}} {2}= \frac 1 {2}( \sum_{j=1}^n r_j^+ - \sum_{j=1}^n r_j^-).$
The distribution of the baricenters $B$ over the disorder realization
is given by:
\begin{eqnarray}
P_B(B,n)=   \int \prod_{j=1}^{n}  \theta(r^+_j-1) \theta(r^-_j-1)
\lambda(r^+_j)\lambda(r^-_j)\nonumber \\
\ \delta \left(B- \frac 1 {2}\sum_{j=1}^n r_j^+ +  \frac 1 {2}
\sum_{j=1}^n r_j^- \right) dr_j^+dr_j^-
\label{baricenter_dist}
\end{eqnarray}
where the sum over the positions $r_j^\pm$ has been replaced by an integral
with a cutoff in  $r_j^\pm=1$. In the simpler case
$\alpha>2$,  the process
is diffusive and the probability that a
walker has reached the $n$-th scattering site is
$p(n,t)\sim t^{-1/2} \exp(-n^2 /(Ct))$,
where $C$ is a suitable constant.
We assume that, after averaging over the walk realizations,
the sites between $X_{n}$ and $X_{- n}$
are visited uniformly. In this framework
the average position of the walker in a certain disorder realization
is given by $x_B=\frac {X_n-X_{- n}} {2} = B$.
In particular, the distribution of $x_B$
at time $t$ is:
\begin{equation}
P(x_B,t)\sim t^{-1/2} \int e^{- n^2/ (Ct)} P_B(x_B,n) dn
\label{distxb1}
\end{equation}
where again we replace the sum over $n$ with an integral.
Taking the Fourier transform with respect $x_B$ one obtains
\begin{equation}
\tilde P(k_B,t)\sim t^{-1/2} \int e^{- n^2 /( {Ct})}
\left(\tilde \lambda( k_B/2) \tilde \lambda(- k_B /2) \right)^n  dn
\label{distkb1}
\end{equation}
where $\tilde\lambda(k)$ is the Fourier transform of
$\lambda(r)$.
Since we are interested in the asymptotic behavior at large
$x_B$ (i.e. at small $k_B$), we get, for $\alpha>2$,
$\tilde \lambda(k)\simeq 1 +i k_B \langle r \rangle -k^2 \langle r^2
\rangle/2$ and then $\tilde \lambda(k) \tilde \lambda(- k) \simeq
1-k^2(\langle r^2 \rangle -\langle r \rangle^2) \simeq
\exp(-k^2 \Delta_r^2)$.
Plunging this expression into Eq. (\ref{distkb1}) we have
\begin{equation}
\tilde P(k_B,t)\sim t^{-1/2}  \int e^{- {n^2}/ (Ct)}
e^{- {k_B^2 n \Delta_r^2/4}} dn.
\label{distkb2}
\end{equation}
Eq. (\ref{distkb2}) entails that
$\tilde P(k_B,t)= \tilde f(k_B/t^{1/4})$ or equivalently
$P(k_B,t)= t^{-1/4} f(k_B/t^{1/4})$ where $\tilde f(\cdot)$ and $f(\cdot)$
are suitable scaling function. Notice that the scaling length of the
process grows as $\ell_B(t) \sim t^{1/4}$ \cite{Bouchaud90}.

For $1<\alpha<2$, the same calculation holds
up to Eq. (\ref{distkb1}), as the process is diffusive also in this case.
However, now  the fluctuations of $\lambda(r)$ are diverging,
hence we obtain at small $k$
$\tilde \lambda(k)\simeq 1 +i k_B \langle r \rangle -D_1 |k|^\alpha$
and  $\tilde \lambda(k) \tilde \lambda(- k) \simeq
\exp(-D_1^2 |k|^{2\alpha})$, where $D_1$ is a suitable constant.
Then, analogously to the previous case we have
\begin{equation}
\tilde P(k_B,t)\sim t^{-1/2} \int e^{-{n^2 }/ (Ct)}
e^{- D_1^2 k_B^{2\alpha} n/4 } dn.
\label{distkb4}
\end{equation}
From Eq. (\ref{distkb4}) we get that the
scaling form of the probability distribution is
$\tilde P(k_B,t)= \tilde f(k_B/t^{1/(2 \alpha)})$, i.e.
$P(k_B,t)= t^{-1/(2 \alpha)} f(k_B/t^{1/(2 \alpha)})$
and $\ell_B(t)\sim t^{1/(2\alpha)}$.

Finally, for $\alpha<1$ the process is not diffusive and
$p(n,t)$ is not Gaussian. However, one can estimate how many
different scattering sites the walker encounters in a time $t$. First,
the number of scattering sites within a distance $\ell$ from the starting
point grows as $n\sim\ell^\alpha$ \cite{Beenakker2009}; then, ignoring
rare long jump events, the typical distance
covered by a walker in a time $t$ is $\ell(t) \sim t^{1/(\alpha+1)}$
\cite{Burioni2010a}. Hence we get $n(t)\sim t^{\alpha/(\alpha+1)}$.
Therefore, we expect that $p(n,t)$ satisfies the scaling form
$p(n,t)\sim t^{-\alpha/(\alpha+1)}g(n/t^{\alpha/(\alpha+1)})$ where $g(\cdot)$
is a suitable scaling function. In the approximation
of a uniform exploration of the interval $[X_n X_{-n}]$ we obtain
the  analogous of  Eq. \ref{distxb1}; i.e.
\begin{equation}
P(x_B,t)\sim t^{-\alpha/(1+\alpha)} \int g\left(\frac{n} {t^{\alpha/(1+\alpha)}}\right)
P_B(x_B,n) dn.
\label{distxb2}
\end{equation}
Now the first moment of $\lambda(r)$ diverges and we have
$\tilde \lambda(k)\simeq 1 -D_2 |k|^\alpha$ and
$\tilde \lambda(k) \tilde \lambda(- k) \simeq \exp(-2 D_1 |k|^{\alpha})$.
Analogously to Eq. \ref{distkb4}, we get for the Fourier transform:
\begin{equation}
\tilde P(k_B,t)\sim t^{-\alpha/(1+\alpha)}
\int \left(\frac{n} {t^{\alpha/(1+\alpha)}}\right)
e^{- 2 D_1 n |k_B|^{\alpha} } dn.
\label{distkb5}
\end{equation}
which implies that the scaling form is
$P(x_B,t)= t^{-1/(1+ \alpha)} f(x_B/t^{1/(1+ \alpha)})$
with $\ell_B(t)\sim t^{1/(1+\alpha)}$

Therefore in the different $\alpha$ regimes  the scaling length
$\ell_B(t)$ governing the dynamics of the baricenter $x_B$ is:
\begin{equation}
\ell_B(t) = \begin{cases}
t^{1/4} &\mbox{for }  \alpha>2 \\
t^{1/(2\alpha)} &\mbox{for }   1<\alpha<2 \\
t^{1/(1+\alpha)} & \mbox{for } \ \alpha<1
\end{cases}
\end{equation}
Comparing $\ell_B(t)$  with the scaling length $\ell(t)$ (\ref{ellcom}),
obtained by averaging both over the disorder and the random walks
realizations, we get that
$\ell_B(t)$ is subleading for $\alpha>1$,
while the two scaling lengths are of the same order of magnitude for $\alpha<1$.
This means that, for $\alpha>1$ the fluctuations due
to random walks dynamics are much larger than the fluctuations due to the different
realization of the disorder. The latter, therefore, are very difficult to measure,
for example observing the probability distribution of a walker in a single
disorder realization $p_n(t)$.
On the other hand, for $\alpha<1$ the distribution $p_n(t)$ should display
distortions due to the disorder realization which are of the same magnitude of
 $\ell(t)$ i.e. the typical size of $p_n(t)$ and therefore they should be more
easily observed in experiments.

If the dynamics does
not present strong anomalous features,
the moments of $\langle x_B^q \rangle$ can be directly evaluated as
$\ell_B(t)^q$. However, when the spacing
between the scattering events are characterized by power laws,
jumps much larger then the scaling length are not exponentially suppressed
and this can give rise to a long tail
$h(x_B,t)$ in the distribution $P(x_B,t)$. The tail $h(x_B,t)$
provides a non trivial contribution to the high order moments,
modifying the value of the exponent $\gamma(q)$ and
giving rise to strongly anomalous diffusion. In particular, $\gamma(q)$
 can be evaluated by means of a single long jump approach
\cite{Burioni2010,Burioni2010a}.
We remark that a jump much larger then the scaling
length affects in the same way the process averaged over the disorder
and the single disorder realization. In particular, the probability of a
long jump can be evaluated as
$h(x_B,t)\sim N(t) \lambda(x_B)$, where $\lambda(x_B)$ is the
probability that one of the segment between scatterers is of length $x_B$,
and $N(t)$ is the number of
scatterer visited by the walker in a time $t$. Therefore we obtain
for the single long jump mechanism the same exponents already evaluated in
\cite{Burioni2010,Burioni2010a}. In particular, the contribution to the moment
$\langle x_B^q\rangle$ grows as $t^{q+0.5-\alpha}$ if $\alpha>1$
and $t^{q-\alpha^2/(1+\alpha)}$ if $\alpha<1$.
Comparing these behaviors with the
contribution to $\langle x_B^q\rangle$ obtained in the scaling
approach i.e. $\ell_B(t)^q$
we get the complete picture for $\gamma(q)$
\begin{eqnarray}
& \mbox{for } \alpha>2 &
\gamma(q) = \begin{cases} \frac{q}{4} &\mbox{for }  q<\frac43 \alpha-\frac23 \\
\frac12+q-\alpha & \mbox{otherwise }   \end{cases} \nonumber \\
& \mbox{for } 1<\alpha<2 &
\gamma(q) = \begin{cases} \frac{q}{2\alpha} &\mbox{for }  q<\alpha \\
\frac12+q-\alpha & \mbox{otherwise }   \end{cases} \\
& \mbox{for } \alpha<1 &
\gamma(q) = \begin{cases} \frac{q}{1+\alpha} &\mbox{for }  q<\alpha \\
\frac{q(1+\alpha)-\alpha^2}{1+\alpha} & \mbox{otherwise }   \end{cases} \nonumber
%
 \end{eqnarray}

\begin{figure}[htbp]
\centering
 \subfigure[$\alpha=0.6$]
   {\includegraphics[width=0.35\textwidth]{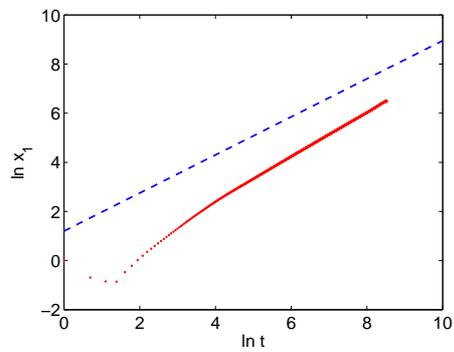}}
 \subfigure[$\alpha=1.5$]
   {\includegraphics[width=0.35\textwidth]{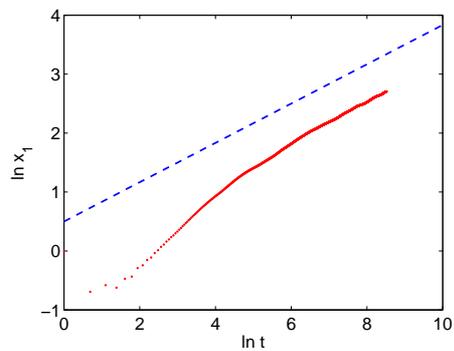}}
\subfigure[$\alpha=3.0$]
   {\includegraphics[width=0.35\textwidth ]{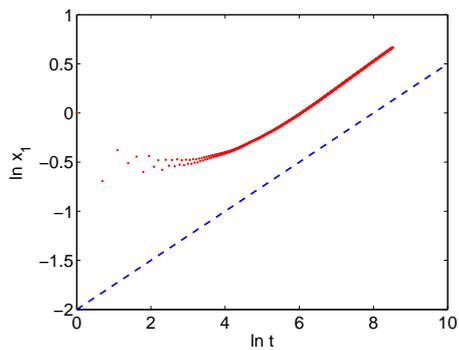}}
 \caption{\protect\footnotesize The moment $x_1$ as defined by (\ref{momentidef}) for three
 values $\alpha$: $\alpha=0.6$ (a), $\alpha=1.5$ (b) e $\alpha=3.0$ (c); averages over $2000$ realizations.
 }
\label{momentoefitalfa06153}
\end{figure}

\begin{figure}[htbp]
\centering
 \subfigure[$\alpha=0.6$]
   {\includegraphics[width=0.35\textwidth]{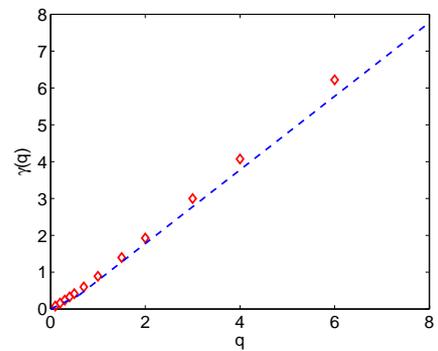}}
 \subfigure[$\alpha=1.5$]
   {\includegraphics[width=0.35\textwidth]{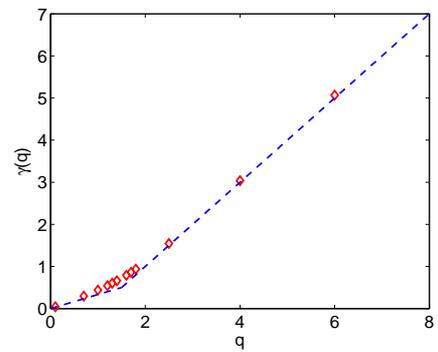}}
\subfigure[$\alpha=3.0$]
   {\includegraphics[width=0.35\textwidth ]{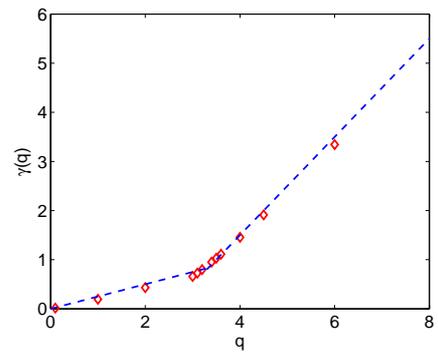}}
 \caption{\protect\footnotesize The scaling exponents $\gamma(q)$ for three
 values of $\alpha$: $\alpha=0.6$ (a), $\alpha=1.5$ (b) e $\alpha=3.0$ (c). The exponent have
 been obtained by power-law fitting of $x_q(t)$; averages are each over a few thousands
 realization of the structure.}
\label{gammaq}
\end{figure}

In fig.~\ref{momentoefitalfa06153} the time-evolution  of $x_1$ in double-logarithmic scale
for the three cases $\alpha=0.6$, $\alpha=1.5$ and $\alpha=3.0$.
In order to avoid any boundary effect, the structures have been generated by
growing two different lattices each of length $N/2$ around the initial site $n_0$
and considering times shorter than $N/2$. Another important limitation for the
numerical test regards the time range where the predicted scaling is expected to
be observable for higher-order moments. To have a sufficient sampling
one needs to consider a number of realizations with a significative
number of jumps much larger of $\ell_B(t)$. For $\alpha>2$,
the probability of obtaining one of this jumps decays with time as
$\int_{\ell_B(t)}^{\infty} h(y,t) dy \sim t^{(2-\alpha)/4}$,
i.e. for $\alpha<2$ the tail $h(x_B,t)$ is subleading with respect to
$P_B(x_B,t)$ for large times.
So the number of realizations needed to observe single long jumps
grows as $t^{(\alpha-2)/4}$  and this means that the
time-range which can be employed to measure the exponents
is bounded from above.
If we take into account this limitation,
we can reliably extract the exponent from the available data.
fig.~\ref{gammaq} shows that, up to the statistical accuracy the
data are in excellent agreement with the analytical estimates.

\section{Conclusions}

We have studied stationary and time-dependent transmission of walkers through finite one-dimensional
lattice where scatterers are arranged on random and deterministic self similar structures with L\'evy distributed disorder.
Using scaling arguments we have predicted the dependence of particle currents on $N$ that
have been successfully tested against the solution of the associated master equation.
The latter approach has considerable advantages with respect to direct Monte Carlo simulations of
the random walk as it allows to obtain averages over trajectories ensembles without
the statistical errors. Altogether the results nicely fit into the global picture that emerged from the recent
literature \cite{Beenakker2009,Burioni2010,Burioni2010a}. Although we focused on one-dimensional
structures only, most of the results carry over to the case of higher-dimensional samples
like $2d$ and $3d$ Le\'vy glasses \cite{Buonsante2011,Groth2012} which are of course 
of major experimental relevance.

A related question we addressed is the problem of the average drift on a specific realization.
We demonstrated that the scaling arguments are of help to understand quantitatively also this
issue. In particular, we showed that when the disorder is characterized by a diverging 
average length of the spacings (i.e. for $\alpha<1$) an anomalous drift arises in the position 
of the baricenter of the probability distribution. Here, the fluctuations due to the different
realization of the disorder  are of the same magnitude of
 $\ell(t)$ i.e. the typical size of $p_n(t)$ and therefore they should be more
easily observed in experiments.

\section*{Acknowledgements}
We  acknowledge useful discussions with R. Livi.


\end{document}